\documentclass[letterpaper]{JHEP}
\usepackage{epsf}

\def\d{\partial}
\def\db{\bar\partial}

\newcommand{\be}{\begin{eqnarray}}
\newcommand{\ee}{\end{eqnarray}}

\newcommand{\beq}{\begin{equation}}
\newcommand{\eeq}[1]{\label{#1}\end{equation}}
\newcommand{\ber}{\begin{eqnarray}}
\newcommand{\eer}[1]{\label{#1}\end{eqnarray}}

\newcommand{\abs}[1]{\left| #1 \right|}
\newcommand{\bra}[1]{\langle{#1}|}
\newcommand{\ket}[1]{|{#1}\rangle}
\newcommand{\ip}[2]{\langle{#1}|{#2}\rangle}
\newcommand{\tr}{\rm{Tr}}
\newcommand{\pl}{\hbox{\small{+}}}
\newcommand{\mi}{\hbox{\hspace{.4mm}\rule[1mm]{1.8mm}{.1mm}\hspace{.4mm}}}

\author{Ulf Lindstr\"{o}m\\ Institute of Theoretical Physics, University
of Stockholm\\ Box 6730\\ S-113 85 Stockholm, SWEDEN\\
\email{ul@physto.se}}
\author{Martin Ro\v{c}ek\\ C.N. Yang Institute of Theoretical Physics,
State University of New York\\ Stony Brook, NY 11794-3840, USA\\
\email{rocek@insti.physics.sunysb.edu}}
\author{Rikard von Unge\\ Department of Theoretical Physics and
Astrophysics\\ Faculty of Science, Masaryk University\\
Kotl\'{a}\v{r}sk\'{a} 2, CZ-611 37, Brno, Czech Republic\\
\email{unge@monoceros.physics.muni.cz}}

\abstract{We study solitons in three dimensional non-commutative scalar
field theory at infinite non-commutativity parameter
$\theta$. We find the metric on the relative moduli space of all solitons
of the form $\ket{n}\bra{n}$ and show that it is K\"{a}hler. We then find
the geodesics of this metric and study the scattering of these solitons.
In particular we find that the scattering is generally right angle for
small values of the impact parameter. We can understand this behaviour in
terms of a conical singularity at the origin of moduli space.}

\title{Non-commutative Soliton Scattering}
\preprint{YITP-00-53}

\begin{document}
\section{Introduction} 
Quantum field theories on non-commutative spaces
have lately received a revival of interest, both as seemingly consistent
non-local deformations of the highly constrained structure of local
quantum field theory, and as theories that appear in various limits of M
theory compactifications \cite{CDS} or the low-energy effective theory of
D-branes in the presence of a background Neveu-Schwarz B-field
\cite{schom,SW}. Conversely, a study of the perturbative properties of
non-commutative field theory reveals that it has many stringy features
\cite{ChRo,Filk,CNPI,CNPII}.

Recently, following the pioneering work \cite{ncsols}, soliton solutions
of non-commutative field theory have been studied
\cite{DMR,HKLM,GN1,Alexios,JMW,GN2,GMS2}. In particular it seems that the
solutions often have nice interpretation as string theory states.
Particularly interesting is the case when the non-commutative solitons are
combined with tachyon condensation \cite{DMR,HKLM,GMS2}. Then one can study
various string states in the standard closed string vacuum without
D-branes. The trick used is to start with a D-brane anti-D-brane pair with
a NS B-field turned on; after the tachyon field condenses into a soliton
configuration, the brane  anti-brane pair annihilate each other so that
only the soliton remains.

This paper is organized as follows. In section \ref{ncslts} we review the
results of \cite{ncsols}. In section \ref{2solsol} we study the simplest
two soliton solution in more detail. In section \ref{simple} we find the
metric on the relative moduli space of the simplest two soliton solution
and find the geodesics to be able to study the scattering. We find that
the metric is K\"{a}hler with a conical singularity in the center which
gives rise to right angle scattering for small impact parameters. In
section \ref{general} we find the metric on the relative moduli space of
soliton solutions of the type $\ket{n}\bra{n}$ for arbitrary n. We give
the
general expression for the metric which is again K\"{a}hler with the same
qualitative behavior for the scattering. Finally we end with a discussion
in section
\ref{discussion}.

\section{Non-commutative solitons}\label{ncslts} 
In this paper we study three dimensional scalar field theory in a space
time where the spacelike coordinates $\hat{x}_{1},\hat{x}_{2}$ are
non-commutative. The energy functional is 
\be\label{Efunc}
 E = \frac{1}{2g^2}\int d^{2}x\left(
  \partial_{t}\phi \partial_{t}  \phi +
 \partial_{1} \phi \partial_{1} \phi +
 \partial_{2} \phi \partial_{2} \phi  + V\left(\phi\right)\right)~,
\ee 
where all fields are multiplied using the non-local star product,
\be
 A\star B = e^{i\frac{\theta}{2}\left(\partial_{1}\partial^{\prime}_{2}
 - \partial_{2}\partial^{\prime}_{1}\right)} A\left(x\right)
B\left(x^{\prime}\right) |_{x=x^{\prime}}~.
\ee 
We are interested in the limit where $\theta\rightarrow\infty$ and to
study that limit we rescale the non-commutative coordinates
$x\rightarrow x\sqrt{\theta}$. Then the star product itself becomes
independent of $\theta$ but we get explicit factors of
$\theta$ in the energy functional in front of the kinetic term and the
potential term $V\left(\phi\right)$. Thus, in the limit
$\theta\rightarrow\infty$, we get the energy functional
\be
  E = \frac{\theta}{2g^2}\int d^{2}x\left(
  \partial_{t}\phi \partial_{t}  \phi + V\left(\phi\right)\right)~.
\ee 
If we are interested in time independent solutions we can drop the
time derivatives and the minimal energy solutions fulfill the equation
\be
 \frac{\partial V}{\partial \phi} = 0~.
\ee 
In \cite{ncsols} it was shown that any field satisfying the relation
\be\label{basicsol}
 \phi \star \phi = \phi~,
\ee 
gives a solution $\lambda \phi$, where $\lambda$ is an extremum of the
{\em function} $V\left(x\right)$. By introducing the creation and
annihilation operators
\be
 a = \frac{\hat{x}_{1} + i \hat{x}_{2}}{\sqrt{2}}~; \;\; 
 a^{\dagger} = \frac{\hat{x}_{1} - i \hat{x}_{2}}{\sqrt{2}}~,
\ee 
they were able to reformulate the problem in terms of the familiar
operators and states of the harmonic oscillator. Namely, any function of
$\hat{x}_{1},\hat{x}_{2}$ can be written as a function of
$a,a^{\dagger}$ which, as an operator, can be rewritten in terms of the
operators 
\be
 \ket{m}\bra{n} = \;\, :\mkern-4mu\frac{a^{\dagger m}}{\sqrt{m!}}
\,e^{-a^{\dagger}a}\frac{a^n}{\sqrt{n!}}\mkern-3mu:~,
\ee 
where the double dots denote normal ordering. In particular, any
function satisfying relation (\ref{basicsol}) is a projection operator and
can alway be written as a sum of the form $\sum_i\,\ket{A_i}\bra{A_i}$ for
arbitrary orthonormal states $\ket{A_i}$. 

Using this method, taking into account that the $\ket{m}\bra{n}$ operators
are normal ordered and the $\phi\left(\hat{x}\right)$ operators are Weyl
ordered, the authors of \cite{ncsols} were able to find a whole set of
radially symmetric solutions $\phi_{n}$ corresponding to the operators
$\ket{n}\bra{n}$,
\be
  \phi_{n} = 2 \left(-1\right)^n e^{-r^2} L_{n}\left(2 r^2\right)~,
\ee 
where $r^2 = x_{1}^{2} + x_{2}^{2}$ and $L_{n}$ are the Laguerre
polynomials. These are all blobs centered at the origin $r =0$.

The energy functional is invariant under unitary transformations of the
operators
\be
\ket{m}\bra{n} \rightarrow U \ket{m}\bra{n} U^{\dagger}~.
\ee 
One particularly interesting unitary operator is
$U=e^{a^{\dagger}z-a\bar{z}}$ which acts as a translation operator. Acting
with this $U$ on any of the above states simply translates the state to be
centered around the point $z =\frac{1}{\sqrt{2}} \left(z_1 + i
z_2\right)$.
We use this in the next section to make solitons that can move around and
scatter off each other.

\section{The two soliton solution}\label{2solsol} 
\begin{figure}[htb]
\begin{center}
 \mbox{\epsfxsize=12cm \epsfbox{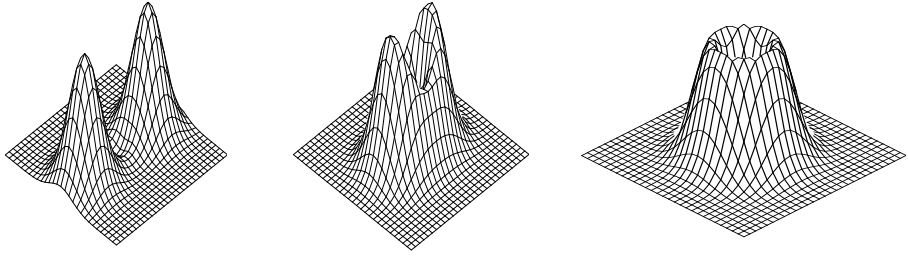}} 
\end{center}
\caption{The 2 soliton solution for $z =$ $2$, $0.9$ and $0$
respectively}\label{ch2fig}
\end{figure}

We want to
study solutions corresponding to two of the basic solitons in the previous
section. By making the positions of these solitons weakly time dependent
(the adiabatic approximation), we can derive a metric
on the relative moduli space and study scattering of these solitons. We
begin with the $\ket{0}\bra{0}$ state. We thus would like to find a
solution to (\ref{basicsol}) which, at large separation of the solitons
should look like
\be
 \phi_{0}\left(x-z\right) + \phi_{0}\left(x+z\right)~,
\ee 
where the relative distance of the solitons is $2z$. From the previous
section we know that
\be
 \phi_{0}\left(x-z\right) \propto U \ket{0}\bra{0} U^{\dagger}~,
\ee 
where $U=e^{a^{\dagger}z-a\bar{z}}$. Since $U$ creates the usual
coherent state $U\ket{0} = e^{-\frac{\abs{z}^2}{2}}e^{a^{\dagger}z}\ket{0}
= \ket{z}$, the solution we are looking for can, for large $z$, be written
as
\be
 \ket{z}\bra{z} + \ket{-z}\bra{-z}~,
\ee 
but unfortunately this is not a good solution for small values of $z$
since the states $\ket{\pm z}$ are not orthogonal. Therefore, the authors
of \cite{ncsols} defined new, mutually orthogonal, states
\be
 \ket{z_{\pm}} = \frac{\ket{z}\pm\ket{-z}}{\sqrt{2\left(1\pm
  e^{-2\abs{z}^2}\right)}}~,
\ee 
in terms of which we can write the solution as
\be
 \lambda \left(\ket{z_+}\bra{z_+} + \ket{z_-}\bra{z_-}\right)~,
\ee
where $\lambda$ is an appropriate normalization. 
Transforming this operator into a wave function we get
\be\label{charge2}
 \frac{\phi_{0}\left(r-z\right) + \phi_{0}\left(r+z\right)
    -2 e^{-2\abs{z}^2}\phi_{0}\left(r\right)\cos\left(2 r \wedge z\right)}
  {1 - e^{-4\abs{z}^2}}~,
\ee 
where $r\wedge z = x_1 z_2 - x_2 z_1$ is a noncommutative factor 
reminiscent of the phases that appear in the star
product. Note that for large separation $z\rightarrow\infty$, the total
wave function becomes that of two $\phi_{0}$ solitons at positions $\pm
z$, whereas when the solitons come very close to each other,
$z\rightarrow 0$, the solution goes to
$2r^2\phi_{0}\left(r\right)$ which does not look like two separate
$\phi_{0}$ solutions, but rather like a new object, a ``charge two''
solution (actually, it is a superposition of $\phi_{1}$ and $\phi_{0}$).
This is exactly what happens for ordinary monopoles \cite{mono}; at large
distance one can see two distinct monopoles but when they get close
together they lose their separate identity and merge into  a charge two
monopole. To illustrate this we have plotted the two soliton solution
(\ref{charge2}) for various values of the separation in figure
\ref{ch2fig}.

\section{Scattering - the simple case}\label{simple} 
Now consider the
two soliton solution (\ref{charge2}) from the previous section and let
the solitons move by making the parameter $z$ time-dependent. If the
solitons move slowly enough, we can make the approximation that the
potential energy of the energy functional (\ref{Efunc}) is always at the
minimum and constant. However, we get a new contribution from the time
derivatives in (\ref{Efunc})\footnote{This analysis is in the spirit of
the discussion of monopole scattering in \cite{manton},
\cite{ah} and and vortex scattering in \cite{Samols92},
\cite{Manton93}. A metric on quantum states was introduces in
\cite{Provost80}.}
\be
 E = \frac{\theta}{g^{2}} \int d^{2}x \left( 2 \partial_{z} \phi
 \partial_{\bar{z}} \phi \dot{z}\dot{\bar{z}} +
 \partial_{z} \phi \partial_{z} \phi \dot{z}\dot{z} +
 \partial_{\bar{z}} \phi \partial_{\bar{z}} \phi \dot{\bar{z}}
\dot{\bar{z}} \right)~.
\ee 
The coefficients of the time derivatives of the $z$ parameters are the
components of the metric of the relative moduli space and the equation of
motion {\em for the $z$'s} is the geodesic equation in this metric. We
thus find that the metric on the relative moduli space of these solitons
is (up to a factor)
\be
 g_{z\bar{z}} &=& \int d^{2}x \partial_{z} \phi\partial_{\bar{z}} \phi~,
 \nonumber\\
 g_{zz} &=& \int d^{2}x \partial_{z} \phi \partial_{z} \phi~, \\
 g_{\bar{z}\bar{z}} &=& \int d^{2}x \partial_{\bar{z}} \phi
 \partial_{\bar{z}} \phi~. \nonumber
\ee 
It is now straightforward but somewhat tedious to plug in the function
(\ref{charge2}) for $\phi$ in these expressions and do the integrals to
find the metric. We find a K\"{a}hler metric with K\"{a}hler potential
$K\left(z,\bar{z}\right) = \ln\sinh\left(2z\bar{z}\right)$.

In polar coordinates $\sqrt{2}z=r e^{i\theta}$ the metric is
$ds^2 = f\left(r\right)\left( dr^2 + r^2 d\theta^2 \right)$
where
\be\label{themet} 
f = \coth\left(r^2\right) -
\frac{r^2}{\sinh^2\left(r^2\right)}~.
\ee 
For large $r$, $f=1$ and the metric becomes flat. This
is expected, since the solitons don't feel each other
when the separation is large, and hence the geodesics are just straight
lines. For small $r$, the function $f$ goes to zero as $r^2$.
Introducing a new coordinate $\rho = r^2$ we obtain
\be
 ds^2 \propto r^2\left( dr^2 + r^2 d\theta^2 \right) =
        \frac{1}{4} \left( d\rho^2 + 4 \rho^2 d\theta^2 \right) =
        \frac{1}{4} \left( d\rho^2 + \rho^2 d\tilde{\theta}^2 \right)~.
\ee 
The last form of the metric looks flat, but the new
coordinate $\tilde{\theta} = 2 \theta$ takes values between $0$ and
$4\pi$, 
and hence there is a conical singularity at the center of the moduli
space.\footnote{The metric (\ref{themet}) has appeared in a different
physical context in \cite{hill,hill2}; however, in that context the
physics
imposed different boundary conditions, and the space of \cite{hill,hill2}
is a $Z_2$ orbifold of our space, and consequently is regular at $r=0$ but
approaches a cone with positive defect $\pi$ for $r\to\infty$.}  
For a cone, it is easy to compute the geodesics:  in the
$(\rho,\tilde{\theta})$ coordinates, the geodesics are straight lines, and
thus a line coming in from $\tilde{\theta}=0$ goes out at
$\tilde{\theta}=\pi$. Consequently, in the physical coordinates
$(r,\theta)$ a geodesic coming in from $\theta =0 $ goes
out at $\theta = \frac{\pi}{2}$, and {\em all} scattering is right
angle, independent of the impact parameter. In our case we expect to see
this behavior only for geodesics that pass close to the singularity,
corresponding to right angle scattering for small impact parameter (just
as in the monopole case), whereas for large impact parameter we expect no
scattering at all since the metric is flat for large separations as
discussed above. In between we expect to see crossover behavior; we have
checked our general arguments numerically.

We can solve the geodesic equation for this metric. The equation can be
integrated to give
\be
\frac{d\theta}{dr} = \pm \frac{1}{r\sqrt{\left(\frac{r}{b}\right)^2 f(r)
 -1}}~,
\ee 
where $b$ is an integration constant. The geodesic can then be
found numerically as
\be
 \theta\left(r\right) = - \int_{\infty}^{r} \frac{ds}
  {s\sqrt{\left(\frac{s}{b}\right)^2 f(s) -1}}~.
\ee 
For large $r$ (where $f\approx 1$) the integral is elementary, and we find
$\theta\left(r\right) \approx \frac{b}{r}$, which implies that $b$
is the impact parameter. The exit angle (if the soliton 
comes in from $\theta=0$) can similarly be found as
\be
 \theta_{\rm{exit}} = - 2 \int_{\infty}^{r_{0}} \frac{ds}
  {s\sqrt{\left(\frac{s}{b}\right)^2 f(s) -1}}~,
\ee 
where we have introduced the constant $r_{0}$, the point of closest
approach; this is related to the impact parameter by $b =
r_{0}\sqrt{f\left(r_{0}\right)}$.

These formulas confirm our expectations about the
scattering behavior. 

\section{Scattering - the general case}\label{general} 
The extremely
simple form of the moduli space metric in the previous section suggests
that there should be a simple way to derive it and to show that it is
K\"{a}hler. To show this we go over to the operator language, and use the
correspondence
\be
 \int d^{2}x \;\d_{t}\phi \d_{t}\phi \leftrightarrow
 \tr \left( \dot{\cal O} \dot{\cal O} \right)~.
\ee
We will perform the derivation for the general case of scattering
between solitons corresponding to the states $\ket{n}\bra{n}$. In
analogy with the previous section we define states $\ket{n,\pm}$ such
that the two soliton solution $\ket{n,+}\bra{n,+}+\ket{n,-}\bra{n,-}$
at large separation look like two separated $\ket{n}\bra{n}$ solitons.
The $\ket{n,\pm}$ state will contain a piece proportional to
$U\left(z\right)\ket{n} \pm U\left(-z\right)\ket{n}$ plus terms
proportional to $\ket{k},\;k<n$ which go to zero exponentially as
$z\rightarrow\infty$.
Thus, the operators we need are always of the form
$\ket{\pl}\bra{\pl} \,+\,
\ket{\mi}\bra{\mi}$ with $\ip{\pl}{\mi} = 0$ and $\ip{\pl}{\pl} =
\ip{\mi}{\mi} = 1$. In terms of these operators, the kinetic 
term from which we read off the metric is
\be
 \tr \left(\left(\,\ket{\dot{\pl}}\bra{\pl} \,+\, \ket{\pl}\bra{\dot{\pl}}
    \, +\,\ket{\dot{\mi}}\bra{\mi} \,+ \,\ket{\mi}\bra{\dot{\mi}}
\,\right)^2
\right)~,
\ee 
which we can rewrite as 
\be
2\left(\,\ip{\dot{\pl}}{\dot{\pl}} \,-\,
\ip{\dot{\pl}}{\pl}\ip{\pl}{\dot{\pl}}
\, -\, \ip{\dot{\pl}}{\mi}\ip{\mi}{\dot{\pl}} 
+\,\ip{\dot{\mi}}{\dot{\mi}} \,-\,
\ip{\dot{\mi}}{\pl}\ip{\pl}{\dot{\mi}}
 \,-\, \ip{\dot{\mi}}{\mi}\ip{\mi}{\dot{\mi}} \,\right)~,
\nonumber\\ 
\ee
since for any {\em constant} $\ip{A}{B}$ we can 
``partially integrate'' $\ip{A}{\dot{B}} = - \ip{\dot{A}}{B}$.  
As explained in \cite{hill,hill2}, this can be rewritten in a more
compact form in terms of the projected derivatives
\be
 \ket{D_{t}\pm} &\equiv& \dot{\mkern2mu\ket{\pm}} \,-\,
\ket{\pm}\langle{\pm}\dot{\mkern2mu\ket{\pm}} 
          \,-\, \ket{\mp}\langle{\mp}\dot{\mkern2mu\ket{\pm}}~,
\ee 
as
\be\label{kin} 
\ip{D_{t}\pl}{D_{t}\pl} \,+\, \ip{D_{t}\mi}{D_{t}\mi}~.
\ee 
Furthermore, in our case, we find that $\ip{\pl}{\dot{\mi}} =
\ip{\mi}{\dot{\pl}} =0$, which makes it possible to write the projected
derivatives in a form that does not mix the $\ket{\pl}$ and $\ket{\mi}$
states
\be\label{dt}
 \ket{D_{t}\pm} &=& \dot{\mkern2mu\ket{\pm}} \,-\,
\ket{\pm}\langle{\pm}\dot{\mkern2mu\ket{\pm}} ~ ,
\ee 
and allows us to calculate the contributions from the $\pm$
states independently.

We calculate the metric at level $n$ by a recursive/inductive
procedure. We consider a state $\ket{n,\pm}$ that is normalized and
orthogonal to all $\ket{k,\pm},\ket{k,\mp}, k<n$, as well as to
$\ket{n,\mp}$.  The state has the form
\beq
\ket{n,\pm}=N_{n,\pm}[a^\dag\frac{\ket{n-1,\pm}} {N_{n-1,\pm}}
+...]=N_{n,\pm}[(a^\dag)^n\frac{\ket{0,\pm}} {N_{0,\pm}} +...]
\eeq{npmdef} 
where $\ket{0,\pm}\equiv N_{0,\pm}(e^{za^\dag}\pm e^{-za^\dag})\ket{0}$
and the terms indicated by $...$ are determined by the orthogonality
conditions; we do not need their explicit form\footnote{We have
choosen the phase such that $N_{n,\pm}$ is always real.}. The metric at
this level is computed from the $z$ and $\bar z$ derivatives of
$\ket{n,\pm}/{N_{n,\pm}}$:
\ber
\ket{\d n,\pm}
\equiv N_{n,\pm}\d\left(\frac{\ket{n,\pm}}{N_{n,\pm}}\right)
\label{Adef}\\ \nonumber\\
\ket{\db n,\pm}\equiv N_{n,\pm}\bar\d
\left(\frac{\ket{n,\pm}}{N_{n,\pm}}\right)
\eer{Bdef} 
The metric depends only on the projections of
$\ket{\d n,\pm}$ and $\ket{\db n,\pm}$ orthogonal to
$\ket{n,\pm}$:
\beq
\ket{D_z n,\pm}\equiv
\ket{\d n,\pm}-\ket{n,\pm}\ip{n,\pm}{\d n,\pm}~,
\eeq{Aperp} 
and similarly for $\ket{\db n,\pm}$; we shall shortly see that
$\ket{\db n,\pm}$ is already orthogonal to $\ket{n,\pm}$, and hence
$\ket{D_{\bar z} n,\pm}=\ket{\db n,\pm}$. We calculate the metric using
\ber
\frac\d{\d t}\ket{n,\pm}&=&\left(\frac\d{\d
t}N_{n,\pm}\right)\left(\frac{\ket{n,\pm}}{N_{n,\pm}}\right)+
\dot z \ket{\d n,\pm}+\dot{\bar z} \ket{\db
n,\pm}~~\Rightarrow\nonumber\\ \nonumber\\
\ket{D_t n,\pm}&=&\dot z \ket{D_z n,\pm} + 
\dot{\bar z}\ket{D_{\bar z} n,\pm}
\eer{dtn}
and substituting into (\ref{kin}).
Since, as we shall see, 
\beq
\ip{\d n,\pm}{D_z n,\pm}=\ip{D_{\bar z}
n,\pm}{\db n,\pm}=0~,
\eeq{ortho}  
the metric is hermitean and takes the form
\beq
g_{z\bar{z}} = G_{n+} + G_{n-},
\eeq{nmetric}
where
\beq 
G_{n,\pm}=~\ip{D_{\bar z} n,\pm}{D_z n,\pm} +\ip{\d n,\pm}{\db n,\pm}~.
\eeq{G} 
For future reference, we also define
\beq 
g_{n,\pm}\equiv ~\ip{D_{\bar z} n,\pm}{D_z n,\pm}~.
\eeq{Anorm} 
We now determine $\ket{D_z n,\pm}$ and $\ket{\db n,\pm}$.  We
begin with the orthogonality relations
\ber
\ip{k,\pm}{n,\pm}&=&0~,~~k<n~,\nonumber \\
\ip{k,\mp}{n,\pm}&=&0~,~~k\le n~,
\eer{ortho2} 
and differentiate with
respect to $z$ and $\bar z$; we find
\ber
\ip{\d k\pm}{n,\pm}+\ip{k,\pm}{\d n,\pm}=0~,&
\nonumber \\
\ip{\d k\mp}{n,\pm}+\ip{k,\mp}{\d n,\pm}=0~,&
\eer{Aeq}
\ber
\ip{\db k\pm}{n,\pm}+\ip{k,\pm}{\db n,\pm}=0~,& 
\nonumber \\
\ip{\db k\mp}{n,\pm}+\ip{k,\mp}{\db n,\pm}=0~.& 
\eer{Beq} 
Our inductive assumption is
\beq
\ket{\db k\pm}\propto\ket{k-1,\mp}~,~k<n~. 
\eeq{induc}
Then from eq.~(\ref{Aeq}) we find that
$\ket{\d n,\pm}$ is orthogonal to all
$\ket{k,\pm},~k<n$ and $\ket{k,\mp},~k\le n$. Consequently, $\ket{\d
n,\pm}$ is a linear combination of $\ket{n+1,\mp}$ and $\ket{n,\pm}$.
This allows us to use eq.~(\ref{Beq}) to determine $\ket{\db n,\pm}$:
\beq
\ket{\db n,\pm}=-\ip{\db {n-1},\mp}{n,\pm}\ket{n-1,\mp}~.
\eeq{Bsol1}
Because $\ket{\d n,\pm}$ is a linear combination of $\ket{n+1,\mp}$ and
$\ket{n,\pm}$, by construction,
$\ket{D_z n,\pm}$ is proportional to
$\ket{n+1,\mp}$. This implies
\ber
\ket{\db n,\pm}
&=&-~\ip{D_{\bar z} {n-1},\mp}{n,\pm}\ket{n-1,\mp}\nonumber\\
&=&-\sqrt{g_{n-1,\mp}}\,\ket{n-1,\mp}~,
\eer{Bsol} 
where the last identity follows from eq.~(\ref{Anorm}). Hence
we have calculated part of the metric (\ref{G}):
\beq
\ip{\d n,\pm}{\db n,\pm}=g_{n-1,\mp}~,
\eeq{Bnorm} 
and proven the inductive assumption (\ref{induc}) holds for $k=n$. We have
also proven that $\bra{n,\mp}\frac\d{\d t}\ket{n,\pm}=0$, as required by
(\ref{dt}), as well as the orthogonality constraints
(\ref{ortho}).
We now calculate  the norm of $\ket{D_z n,\pm}$ by differentiating
$\ip{n,\pm}{n,\pm}=1$ with respect to $z$; using the definitions of the
states $\ket{\d n,\pm},\ket{\db n,\pm}$ (\ref{Adef},\ref{Bdef}), and the
orthogonality relation $\ip{\d n,\pm}{n,\pm}=0$, we find
\beq
\ip{n,\pm}{\d n,\pm}~+~2\d\ln(N_{n,\pm})~=~0~.
\eeq{nA} 
We may differentiate this relation with respect to $\bar z$ to
obtain:
\ber
\ip{\db n,\pm}{\d n,\pm} &+&2\bar\d\ln(N_{n,\pm})
\ip{n,\pm}{\d n,\pm}\nonumber\\&+&
\bra{n,\pm}N_{n,\pm}\bar\d
\left(\frac{\ket{\d n,\pm}}{N_{n,\pm}}\right)
~+~2\d\bar\d\ln(N_{n,\pm})~=~0~.
\eer{AA}  
Substituting (\ref{nA}) into (\ref{AA}) gives
\beq 
~\ip{D_{\bar z} n,\pm}{D_z n,\pm} ~=~-\bra{n,\pm}N_{n,\pm}\bar\d
\left(\frac{\ket{\d n,\pm}}{N_{n,\pm}}\right) ~-~2\d\bar\d\ln(N_{n,\pm})~.
\eeq{ApAp} 
Differentiating the orthogonality relation
$\ip{n,\pm}{\db n,\pm}=0$, we find
\beq
\ip{\d n,\pm}{\db n,\pm} ~+~\bra{n,\pm}N_{n,\pm}\d
\left(\frac{\ket{\db n,\pm}}{N_{n,\pm}}\right)~=~0~;
\eeq{BB} 
by the definitions of 
$\ket{\d n,\pm},\ket{\db n,\pm}$ (\ref{Adef},\ref{Bdef}), we have
\beq
\d\left(\frac{\ket{\db n,\pm}}{N_{n,\pm}}\right)~=~
\bar\d
\left(\frac{\ket{\d n,\pm}}{N_{n,\pm}}\right)~,
\eeq{AB} 
and hence the norm of $\ket{D_z n,\pm}$ is (the square root of)
\beq 
g_{n,\pm}~=~\ip{\d n,\pm}{\db n,\pm} ~-~2\d\bar\d\ln(N_{n,\pm})~.
\eeq{gf} 
Consequently,
\beq 
g_{n,\pm}=-\d\bar\d
\ln(N^2_{n,\pm})+g_{n-1,\mp}=-\d\bar\d
\ln(N^2_{n,\pm}N^2_{n-1,\mp}...N^2_{0,\pm(-1)^n}) ~, 
\eeq{kahler} 
and the full metric obeys
\beq 
G_{n,\pm}=-\d\bar\d
\ln(N^2_{n,\pm})+2g_{n-1,\mp}~. 
\eeq{Gkahler}
Clearly, this gives us an expression for the K\"ahler potential in terms
of $N^2_{k,\pm}$ for $k\le n$. 

To complete the determination of the metric, we need a formula
for $N^2_{n,\pm}$; we calculate it, and find a second expression
for the metric.  From the definition of $\ket{\d n,\pm}$ (\ref{Adef})
and the orthogonality properties found above, we have
\beq
\ket{D_z n,\pm}~=~
\frac{N_{n,\pm}}{N_{n,\mp}}\left(a^\dag\ket{n,\mp}-
\ket{n,\pm}\bra{n,\pm}a^\dag\ket{n,\mp}\right)~.
\eeq{Aperp2} 
To evaluate the second term, we need to find
$a\ket{n,\pm}$. The leading term follows from the definition of
$\ket{n,\pm}$ in eq.~(\ref{npmdef}), and we are led to the ansatz
\beq 
a\ket{n,\pm}=z\frac{N_{n,\pm}}{N_{n,\mp}}\ket{n,\mp}+
C_{n,\pm}\ket{n-1,\pm}+...
\eeq{aonnpm} 
where the terms represented by $...$ involve
$\ket{k,\pm},\ket{k,\mp}$ for $k<n-1$ and will be shortly shown to vanish.
Consider the inner products
\beq
\bra{k,\pm}a\ket{n,\pm}~,~~\bra{k,\mp}a\ket{n,\pm}~;
\eeq{inner} 
for $k<n-1$, since $a^\dag\ket{k,\pm}\,\propto
\ket{k+1,\pm}+...$, the orthogonality properties of $\ket{n,\pm}$
immediately imply that all these inner products vanish. Hence, as promised
above, the missing terms in (\ref{aonnpm}) do indeed vanish. For $k=n-1$, 
we use
\beq 
a^\dag\ket{n-1,\pm}=
\frac{N_{n-1,\pm}}{N_{n,\pm}}\ket{n,\pm} +...
\eeq{adagn} 
Substituting this into (\ref{aonnpm},\ref{inner}), we find
\beq 
C_{n,\pm}=\frac{N_{n-1,\pm}}{N_{n,\pm}}~,
\eeq{c} 
and hence
\beq 
a\ket{n,\pm}=z\frac{N_{n,\pm}}{N_{n,\mp}}\ket{n,\mp}+
\frac{N_{n-1,\pm}}{N_{n,\pm}}\ket{n-1,\pm}~.
\eeq{aonnpmf}
We thus find an explicit expression for 
$\ket{D_z n,\pm}$:
\beq
\ket{D_z n,\pm}~=~
\frac{N_{n,\pm}}{N_{n,\mp}}\left(a^\dag\ket{n,\mp}-
\bar z
\frac{N_{n,\pm}}{N_{n,\mp}}\ket{n,\pm}\right)~.
\eeq{Aperp3} 
Because
$\ket{n,\pm}\,\propto\ket{D_z {n-1}\mp}$, we find a recursive relation for
the normalization factor $N_{n,\pm}$:
\beq 
N_{n,\pm}=\frac{N_{n-1,\mp}}{\sqrt{g_{n-1,\mp}}}~,
\eeq{Nnorm}
which allows us to calculate the metric recursively using (\ref{kahler}).
Alternatively, we can calculate the contribution (\ref{Anorm}) to the
metric directly from (\ref{Aperp3}):
\beq 
g_{n,\pm}=
\left(\frac{N_{n,\pm}}{N_{n,\mp}}\right)^2
\left[1+\left(\frac{N_{n-1,\mp}}{N_{n,\mp}}\right)^2
\right]+z\bar z
\left[1-\left(\frac{N_{n,\pm}}{N_{n,\mp}}\right)^4\right]~.
\eeq{gpm} 
This bears no obvious resemblance to (\ref{kahler}); we have
explicitly verified the consistency of the formulas (\ref{gpm}) and
(\ref{kahler}) to level $n=3$. 

Though the metrics increase in complexity as $n$ increases, their
qualitative behavior for large and small $r$ is the same, and
consequently,
the scattering of higher $n$ solitons should be basically the same as for
the $n=0$ case discussed in the previous section.

\section{Discussion}\label{discussion} 
In \cite{hill} and \cite{hill2} it is described
how the assumption that the state vectors are
analytic functions of the complex moduli space coordinates, up
to normalization factors, leads to a K\"{a}hler
metric.  Although this assumption does not hold for our case in
this paper, we have shown that the metric on the relative moduli
space of two solitons of the type
$\ket{n}\bra{n}$ for arbitrary $n$ is nevertheless K\"{a}hler,
and we have given a general
expression for the K\"{a}hler potential of this metric.
Our analysis was done for the
case where the non-commutativity parameter
$\theta\rightarrow\infty$. By finding the geodesics of this metric we
studied the scattering of these solitons against each other and found a
universal behavior in that the scattering always goes to right angle
scattering for small values of the impact parameter. This we understood as
coming from a conical singularity at the center of the moduli space.

A natural and important question is what happens at finite
$\theta$. We hope that the methods that we
have developed in this paper can also be used in this case. Another
important question where we should be able to use our methods is in
theories with more non-commutative coordinates. One could also imagine
studying nonspherically symmetric solutions by choosing a different
$U$ operator corresponding to squeezed states.

An equally interesting question is to take the viewpoint of
\cite{DMR,HKLM} where it was shown that the solitons treated in this paper can
be seen as lower dimensional D-branes of bosonic string theory. Following
this logic what we have done in this paper is to study D-brane scattering.
It would be interesting to compare our results with more direct
calculations of these processes in the D-brane language.

It would also be interesting to investigate scattering of soliton
solutions of other type of non-commutative theories. In particular it
should be interesting to study the soliton solutions of non-commutative
Yang-Mills theory in this context.\\ 
 
\noindent{\bf Note added}(in response to a question raised by the
referee):
\medskip
 
\noindent In this paper we consider only moduli
corresponding to translations. The symmetry group of the solutions is
$U(\infty )$; this means that the moduli space is infinite
dimensional. The motivation for our choice comes from considerations at
finite $\theta$, where the degeneracy along almost all of the moduli is
lifted.  Following \cite{ncsols} we can write the derivative terms that we
dropped at infinite $\theta$ as
\be
\frac{2\pi}{g^2}Tr \left([a,A]
[A,a^{\dagger}]\right)~,
\ee
where $A$ is an operator corresponding to the state in question. 
The contribution to the energy will in general depend on the moduli parameters
and constitutes a potential for them. In the one soliton case,
$A = U \ket{n}\bra{n} U^{\dagger}$, the $U$ that leads to a minimal (constant)
contribution is a translation. It has (additional) energy
\be
\frac{2\pi}{g^2}(2n+1)~,
\ee
{\it i.e.}, $2{g^2}/\pi$ for the lowest $n=0$
state. In the two soliton sector, where $A = \ket{n+}\bra{n+} +
\ket{n-}\bra{n-}$, we again find independence of the moduli parameter. The
energy contribution is 
\be
\frac{2\pi}{g^2}2(2n+1)~,
\ee
{\it i.e.}, $4{g^2}/\pi$ for the lowest $n=0$ state. Thus, the moduli
spaces that we study for the lowest level remain degenerate to first order in
perturbation theory, whereas the degeneracy is lifted for generic moduli. 
These and other aspects on multisoliton solutions will be discussed in a
forthcoming paper \cite{multisol}.

\acknowledgments{We are happy to thank L. Hadasz for comments and discussions.
MR is happy to thank UL, RvU and the Department of Theoretical Physics and
Astrophysics at the Faculty of Science, Masaryk University, for their
hospitality. The work of UL was supported in part by NFR grant 650-1998368
and by EU contract HPRN-CT-2000-0122. The work of RvU was supported by the
Czech Ministry of Education under Contract No. 144310006. The work of
MR was supported in part by NSF grant PHY9722101.}


\begin{thebibliography}{666}
\bibitem{CDS}A.~Connes, M.~Douglas and A.~Schwarz, ``Noncommutative
Geometry and Matrix Theory: Compactification on Tori''
\jhep{02}{1998}{003}, {\tt hep-th/9711162}.

\bibitem{schom}V.~Schomerus, ``D-branes and Deformation Quantization'',
\jhep{06}{1999}{030}, {\tt hep-th/9903205}.

\bibitem{SW}N.~Seiberg and E.~Witten, ``String Theory and Noncommutative
Geometry'', \jhep{09}{1999}{032}, {\tt hep-th/9908142}.

\bibitem{ChRo}I.~Chepelev and R.~Roiban, ``Renormalization of Quantum
Field Theories on Noncommutative $R^{D}$, I. Scalars'',
\jhep{05}{2000}{037}, {\tt hep-th/9911098}.

\bibitem{Filk}T.~Filk, ``Divergencies in a Field Theory on Quantum
Space'', \plb{376}{1996}{53}.

\bibitem{CNPI}S.~Minwalla, M.~van~Raamsdonk and N.~Seiberg,
``Noncommutative Perturbative Dynamics'', {\tt hep-th/9912072}.

\bibitem{CNPII}M.~van~Raamsdonk and N.~Seiberg, ``Comments on
Noncommutative Perturbative Dynamics'', \jhep{03}{2000}{035}, {\tt
hep-th/0002186}.

\bibitem{ncsols}R.~Gopakumar, S.~Minwalla and A.~Strominger,
``Noncommutative Solitons'', \jhep{05}{2000}{020}, {\tt hep-th/0003160}.

\bibitem{DMR}K.~Dasgupta, S.~Mukhi and G.~Rajesh, ``Noncommutative
Tachyons'', \jhep{06}{2000}{022}, {\tt hep-th/0005006}.

\bibitem{HKLM}J.~A.~Harvey, P.~Kraus, F.~Larsen and E.~J.~Martinec,
``D-branes and Strings as Non-commutative Solitons'',
\jhep{07}{2000}{042}, {\tt hep-th/0005031}.

\bibitem{GN1}D.~J.~Gross and N.~A.~Nekrasov, ``Monopoles and Strings in
Noncommutative Gauge Theory'', \jhep{07}{2000}{034}, {\tt hep-th/0005204}.

\bibitem{Alexios}A.~P.~Polychronakos, ``Flux tube solutions in
noncommutative gauge theories'', {\tt hep-th/0007043}.

\bibitem{JMW}D.~P.~Jatkar, G.~Mandal and S.~R.~Wadia, ``Nielsen-Olesen
Vortices in Noncommutative Abelian Higgs Model'', {\tt hep-th/0007078}.

\bibitem{GN2}D.~J.~Gross and N.~A.~Nekrasov, ``Dynamics of Strings in
Noncommutative Gauge Theory'', {\tt hep-th/0007204}.

\bibitem{GMS2}R.~Gopakumar, S.~Minwalla and A.~Strominger, ``Symmetry
Restoration and Tachyon Condensation in Open String Theory'', {\tt
hep-th/0007226}.

\bibitem{mono}R.S.~Ward,
``A Yang-Mills-Higgs monopole of charge 2'',
\cmp{79}{1981}{317}.

\bibitem{Provost80}J.P. Provost and G. Vallee,
{\em Riemannian Structure on Manifolds of Quantum States},
\cmp{76}{1980}{289}. 

\bibitem{manton}N.S.~Manton,
``A remark on the scattering of BPS monopoles'',
\plb{110}{1982}{54}. 

\bibitem{ah}M.F.~Atiyah and N.J.~Hitchin,
``Low energy scattering of nonabelian monopoles'',
\pla{107}{1985}{21}. 

\bibitem{Samols92} T.M. Samols, {\em Vortex Scattering},
\cmp{145}{1992}{149}.

\bibitem{Manton93} N.S. Manton, {\em Statistical mechanics of
vortices},
\npb{400}{1993}{624}.

\bibitem{hill}T.H.~Hansson, S.B.~Isakov, J.M.~Leinaas, U.~Lindstr\"om,
``Classical phase space and statistical mechanics of identical
particles'',
{\tt quant-ph/0003121}.

\bibitem{hill2}T.H.~Hansson, S.B.~Isakov, J.M.~Leinaas, U.~Lindstr\"om,
``Exclusion Statistics in Classical Mechanics',
{\tt quant-ph/0004108}.

\bibitem{multisol}
 L. Hadasz, U. Lindstr\"om, M. Ro\v cek, R. v. Unge, in preparation.

\end{thebibliography}
\end{document}